\documentclass[aps,prl,preprintnumbers,groupedaddress]{revtex4}
\newcommand{\bea}{\begin{eqnarray}}
\newcommand{\eea}{\end{eqnarray}}
\usepackage{graphicx}
\begin{document}
\preprint{nucl-th/0410014v4}
\title{A Pad\'e-aided analysis \\ of nonperturbative $NN$ scattering in $^1S_0$ channel }
\author{Ji-Feng Yang, Jian-Hua Huang}
\address{Department of Physics, East China Normal University,
Shanghai 200062, China}

\begin{abstract} We carried out a Pad\'e approximant
analysis on a compact factor of the $T$-matrix for $NN$ scattering
to explore the nonperturbative renormalization prescription in a
universal manner. The utilities and virtues for this Pad\'e
analysis were discussed.
\end{abstract}
\maketitle
\section{ Introduction}
Since Weinberg's seminal work\cite{WeinEFT}, the effective field
theory (EFT) approach to the nuclear forces has been extensively
investigated\cite{BevK}. However, such applications are plagued by
severe nonperturbative UV divergence. For the EFT approach to be
useful, the regularization and subtraction scheme must be
carefully worked out together with a consistent set of power
counting rules. In other words, appropriate renormalization
prescription is needed in nonperturbative regimes. There have been
many contributions to this
issue\cite{vK,Epel,Steele,Rho,KSW,Fred,Gege,Soto,BBSvK,RGE,Oller,Nieves,VA,Higa}
(More could be found in Ref.\cite{BevK}), among which there are
some controversies and debates. At some points, different
approaches could lead to rather disparate
predictions\cite{chLnforce}.

Recently, a compact parametrization of the $T$-matrix is proposed
in Ref.\cite{g_npt}, with which the obstacle for renormalization
being identified as the compact form of the $T$-matrix. In a
concrete example\cite{contact}, it was shown that the $T$-matrix
could only be renormalized through 'endogenous' counter terms,
which result in nontrivial prescription dependence. Such
prescription dependence must be removed or fixed by imposing
appropriate physical boundary conditions, for instance, through
certain procedure of data fitting, as is frequently done in
literature. The conventional power counting could be preserved
within such procedures.

In this short report, we sketch a Pad\'e approximant analysis
basing on the aforementioned parametrization for the $T$-matrix.
The paper is organized as follows: In Sec. II, the parametrization
proposed in Re.f\cite{g_npt} is briefly described with some
related remarks. In Sec. III, the Pad\'e approximant of a factor
in the compact parametrization of the $T$-matrix is employed to
parametrize the nonperturbative prescription dependence. Then
predictions for phase shifts are made at various chiral orders and
Pad\'e approximants. The instrumental utilities and other aspects
of this analysis will be discussed in Sec. IV. The report is
summarized in Sec. IV.

\section{The compact parametrization}
To describe low energy $NN$ scattering, one first constructs the
potential $V$ from $\chi$PT\cite{WeinEFT} up to certain chiral
order, then computes the $T$-matrix through Lippmann-Schwinger
equation (LSE), which is plagued with severe nonperturbative UV
divergences. To appreciate the crucial aspects, a compact form of
$T$-matrix in diagonal channels\cite{foot} is proposed in
Ref.\cite{g_npt} as follows basing on LSE,
\begin{eqnarray}
\label{npt} &&\frac{1}{T_{^{2S+1}L_J}(p^{\prime},p;
E)}=\frac{1}{V_{^{2S+1}L_J}(p^{\prime},p)}-
{\mathcal{G}}_{^{2S+1}L_J}(p^{\prime},p; E),\\ \label{G}
&&{\mathcal{G}}_{^{2S+1}L_J}(p^{\prime},p; E) \equiv\frac{
\displaystyle\int\frac{kdk^2}{(2\pi)^2}
V_{^{2S+1}L_J}(p^{\prime},k ) G_0(k;E) T_{^{2S+1}L_J}(k,p;
E)}{V_{^{2S+1}L_J}(p^{\prime},p)T_{^{2S+1}L_J}(p^{\prime},p; E)}.
\end{eqnarray}Here $G_0(k;E)\equiv 1/(E-k^2/M+i
\epsilon)$, with $M$ being nucleon mass, $p^{\prime}$ and $p$
being the off-shell external momenta. Using the on-shell relation
between $K$- and $T$-matrix\cite{newton}: $ 1/{T(p)}=1/{K(p)}+i
\frac{M} {4\pi}p$, we arrive at the following on-shell relations
(from now on, we omit the subscript '$^{2S+1}L_J$')
\begin{eqnarray}\label{G_K}{\mathcal{G}}(p)&&=V^{-1}(p)-
K^{-1}(p)- i\frac{M} {4\pi}p;\\ \label{OS}
T^{-1}(p)&&=V^{-1}(p)-{\mathcal{G}}(p).
\end{eqnarray} Obviously, ${\mathcal{G}}$ assumes all the nonperturbative
divergences in a compact form.  Any approximation to the quantity
${\mathcal{G}}$ leads to a nonperturbative scheme for $T$. Here we
should remind that the power counting is applied in the
construction of the potential $V(p,p^{\prime})$.

In perturbation theory, UV divergences are removed order by order
{\em before} the amplitudes are summed up. While for
Eq.(\ref{npt}) or (\ref{OS}) in nonperturbative regime, infinitely
many UV divergent amplitudes like $VG_0VG_0VG_0\cdots V$ must be
lump summed into a compact form. Then one must specify the order
for implementing the following two incommutable procedures:
subtraction versus nonperturbative summation. So a natural
discrimination arises between 'endogenous' and 'exogenous' counter
terms (or equivalent operations) that are introduced {\em before}
and {\em after} the summation respectively\cite{g_npt,contact}.
The compact form of $T$-matrix fails the 'exogenous' counter
terms\cite{g_npt,contact}. In other words, the renormalization
through 'endogenous' counter terms is the only sensible procedure
in nonperturbative regime. Since the Schr\"odinger equation
approach\cite{BBSvK} is intrinsically nonperturbative, any
successful subtraction in the Schr\"odinger equation approach
serves as a concrete instance for 'endogenous' renormalization in
action\cite{g_npt,contact}. In practice, 'endogenous' subtraction
is a formidable task in the $T$-matrix formalism: To complete the
subtraction AND summation to ALL orders! That is why various forms
of finite cutoff prevail in literature. In whatever approaches,
the compact form of $T$-matrix persists and makes nonperturbative
prescription dependence strikingly different from the perturbative
cases\cite{scheme,contact}: Physical boundary conditions must be
imposed as a nontrivial procedure.

In what follows, all the EFT couplings are collectively denoted by
$[C_{\ldots}]$ and pion mass by $m_{\pi}$. In general, any
prescription could be parametrized by a set of dimensionless
constants $[q_{\ldots}]$ and a dimensional scale $\mu$, including
various finite cutoff schemes.

\section{a Pad\'e-aided analysis of the EFT for $NN$ scattering}
\subsection{Motivation}
From the above discussions, it is clear that the nonperturbative
renormalization prescription is solely assumed in the factor
${\mathcal{G}}$. Obviously, the ${\mathcal{G}}$ factor could not
be perturbative in terms of the EFT couplings, and its nontrivial
nonperturbative prescription dependence is parametrized by the
parameters to be physically fixed. These points have been
demonstrated in Ref.\cite{contact}. From the point of view of
Eq.(\ref{npt}), the main issue in the EFT for low energy $NN$
scattering is to work out an appropriate procedure or prescription
of renormalization in the nonperturbative regime so that the EFT
power counting schemes (encoded in $V$) remain intact. This
requires the full solution of the nonperturbative factor
${\mathcal{G}}$.

Now we need a general formalism to describe the nontrivial
features of the renormalization in the nonperturbative regimes.
Due to the difficulty in obtaining the full analytical
nonperturbative solutions, certain approximation must be employed
in practice, including various numerical approaches\cite{Epel}.
Here, we wish to propose an approximation approach that is, we
feel, more analytical in the nonperturbative regime. The starting
point is just the compact parametrization defined in
Eq.(\ref{npt}). We should note that in this parametrization the
direct EFT component is the potential, which is constructed using
EFT power counting. Thus the EFT is naturally incorporated in the
following analysis through the potential. We will return to this
issue in the third subsection.

\subsection{Formulation}
The idea is very simple, we parametrize the factor ${\mathcal{G}}$
in terms of Pad\'e approximant. This is reasonable as
${\mathcal{G}}$ is nonperturbative in terms of the EFT couplings
and prescription parameters. In formulae, we employ the following
parametrization of the ${\mathcal{G}}$ factor ($p=\sqrt{M E}$):
\begin{eqnarray}
\label{fraction} {\mathcal{G}}(p)\|_{\text{Pad\'e}}
\Rightarrow\left \{\frac{\nu_0+\nu_1 p^2
+\cdots}{\delta_0+\delta_1 p^2 +\cdots}-  \frac{M }{4\pi}ip\right
\}\|_{\text{Taylor}}\Rightarrow g_{(0)}+g_{(1)} p^2+\cdots-
\frac{M }{4\pi} ip.
\end{eqnarray}Here, the Taylor series is also listed as an
expansion in much lower energy regions. Note the significant
distinction between the Pad\'e analysis of the ${\mathcal{G}}$
factor here and that of the whole $T$-matrix: In the former case,
EFT is indispensable in the systematic construction of the kernel
(potential), while in the latter case, EFT plays no role at all.

Obviously ${\mathcal{G}}$ assumes all the prescription dependence
through $\nu_i,\delta_j$ or $g_{(n)}$ that are nontrivial
functions of the EFT couplings $[C_{\ldots},M,m_{\pi}]$, and the
prescription parameters, $[q_{\ldots},\mu]$. Then instead of
$[q_{\ldots},\mu]$, we could use $\nu_i,\delta_j$ or $g_{(n)}$ to
parametrize the renormalization prescription of the $T$-matrix
within the EFT approach,\bea\label{padepara_T}
&&T^{-1}(p;[C_{\ldots},M,m_{\pi}];[\nu_{\ldots},\delta_{\ldots}])
\Rightarrow V^{-1}(p;[C_{\ldots},M,m_{\pi}])- \frac{\nu_0+\nu_1
p^2 +\cdots}{\delta_0+\delta_1 p^2 +\cdots}+  \frac{M
}{4\pi}ip,\eea or \bea\label{taylorpara}
&&T^{-1}(p;[C_{\ldots},M,m_{\pi}];[g_{(\ldots)}]) \Rightarrow
V^{-1}(p;[C_{\ldots},M,m_{\pi}])- (g_{(0)}+ g_{(1)} p^2+\cdots)+
\frac{M }{4\pi}ip.\eea

Intuitively, nonperturbative prescription dependence in
$\nu_{i},\delta_{j}$ or $g_{(n)}$ could be understood from the
rigorous nonperturbative solution of on-shell $T$-matrix for
$^1S_0$ channel scattering with contact potential at
next-to-leading order (Nlo) and next-to-next-to-leading order
(Nnlo)\cite{contact}: \bea \text{Nlo}: V_{(2)}&&= C_0 +C_2
(p^2+{p^{\prime}}^2);\nonumber
\\\Rightarrow
 \label{Tn2}
T^{-1}&&=\frac{(1-C_2J_3)^2}{C_0+C_2^2
J_5+C_2(2-C_2J_3)p^2}+ J_0 +\frac{M }{4\pi}ip;\\
\label{localV3}\text{Nnlo}: V_{(4)}&&=C_0 +C_2
(p^2+{p^{\prime}}^2)+\tilde{C}_4p^2{p^{\prime}}^2+C_4
(p^4+{p^{\prime}}^4);\nonumber \\ \Rightarrow \label{Tn3}
T^{-1}&&=\frac {N_0+N_1 p^2+N_2 p^4}{D_0+D_1p^2+D_2p^4+ D_3 p^6}+
J_0 +\frac{M }{4\pi}ip,\eea with $[N_i,D_j]$, which correspond to
$[\nu_{i},\delta_{j}]$, being polynomials in terms of coupling
$[C_{\ldots}]$ and $[J_n]$. Here the constants $[J_n]$ come from
divergent loop integrals and hence parametrize the renormalization
prescription, like the set $[q_{\ldots},\mu]$\cite{contact}. Then
$N_i,D_j$ effectively parametrize a prescription due to their
nontrivial dependence on $[J_{n}]$. In Pad\'e approximant,
$\nu_{i},\delta_{j}$ or $g{(n)}$ take over the role of $N_i,D_j$,
and the physical boundary conditions are to be imposed on
$\nu_{i},\delta_{j}$ or $g_{(n)}$.
\subsection{Power counting and nonperturbative prescription}
As is stressed in the previous section, EFT and its power counting
rules enter through the potential. Through Eq.(\ref{npt}) the EFT
elements and their power counting rules are carried over to the
$T$-matrix in a nonperturbative manner. Since the kernel of the
LSE or the potential is perturbative, its renormalization is still
perturbatively implementable within the EFT power counting rules.
Therefore, for the $T$-matrix, Eq.(\ref{npt}) just provides a
concise separation of the nonperturbative renormalization
information from other things. Both ${\mathcal{G}}$ and $V$ are in
principle EFT objects, the sole and important distinction is that
the former exclusively carries the information about the
nonperturbative renormalization prescription. Thus, we could
append some subscripts to Eq.(\ref{npt}) as
follows,\begin{eqnarray} \label{OS}
T^{-1}_{\text{EFT,n.p.t.}}(p)&&=V^{-1}_{\text{EFT,p.t.}}(p)-
{\mathcal{G}}_{\text{EFT,n.p.t.}}(p).
\end{eqnarray}

In general, within a natural EFT, the nonperturbative objects
(e.g. ${\mathcal{G}}$) should also exhibit certain degree of
naturalness in the sense that, the scales involved in such objects
should not deviate very much from the natural sizes. But the
renormalization in nonperturbative regimes does allow for other
unconventional scenarios, without violating natural power counting
rules\cite{contact}. As a matter of fact, the Pad\'e analysis of
${\mathcal{G}}$ alone according to Eq.(\ref{npt}) does not affect
anything of the EFT power counting rules, i.e., the implementation
of the EFT power counting rules and the nonperturbative
renormalization procedures are disentangled. The possible
subtleness in the nonperturbative factor ${\mathcal{G}}$ should
not be misunderstood as the inconsistency or the unnaturalness of
the EFT power counting rules or even as the inapplicability of EFT
method at all. That is, the unnaturalness in $\nu_{i},\delta_{j}$
or $g_{(n)}$ could have nothing to do with the inconsistency of
EFT power counting.

\subsection{Fitting and predictions: $^1S_0$ channel}
With the preceding preparations, we can demonstrate the
predictions of phase shifts for the $^1S_0$ channel $NN$
scattering at different orders of potential using different Pad\'e
approximants. For each case, the prescription parameters, i.e.,
the Pad\'e parameters are fixed through fitting the phase shifts
in the low energy ends. The laborious loop integrations and
'endogenous' subtractions are naturally avoided.

We will employ the potentials that are worked out in
Ref.\cite{Epel} (denoted as EGM from now on), which contains no
energy dependence, and less contact couplings--a favorable aspect
for fitting the 'physical' values for $\nu_{i},\delta_{j}$ or
$g_{(n)}$. One could well employ other construction schemes for
potentials. In fact, one could compare any pair of potential
schemes only through fitting the parameters $\nu_{i},\delta_{j}$
or $g_{(n)}$. It is obvious that at any chiral order with any
Pad\'e approximant, different Pad\'e parameters would yield rather
different phase shifts curves. We will not show the figures for
demonstrating such nontrivial prescription dependence due to space
limitation. Let us focus on more interesting figures with the
Pad\'e parameters determined through boundary conditions: fitting
in the low energy regions: (1) At Lo, $\mathrm{T}_{\text{lab}}\in
(0,3)$ in units of $\texttt{MeV}$;(2) At Nlo , while
$\mathrm{T}_{\text{lab}}\in (0.2, 13)$; (3) At Nnlo,
$\mathrm{T}_{\text{lab}}\in (3,23)$. The phase shift is obtained
from the following formula,\bea\delta(p)=\arctan \left
\{-\frac{Mp}{4\pi} \left (\frac{1}{V(p)}-
\text{Re}({\mathcal{G}}(p))\right )^{-1}\right \}. \eea

\begin{figure}[h]
\begin{center}
\begin{tabular}{ccc}
\hspace{-0.3cm} \resizebox{60mm}{!}{\includegraphics{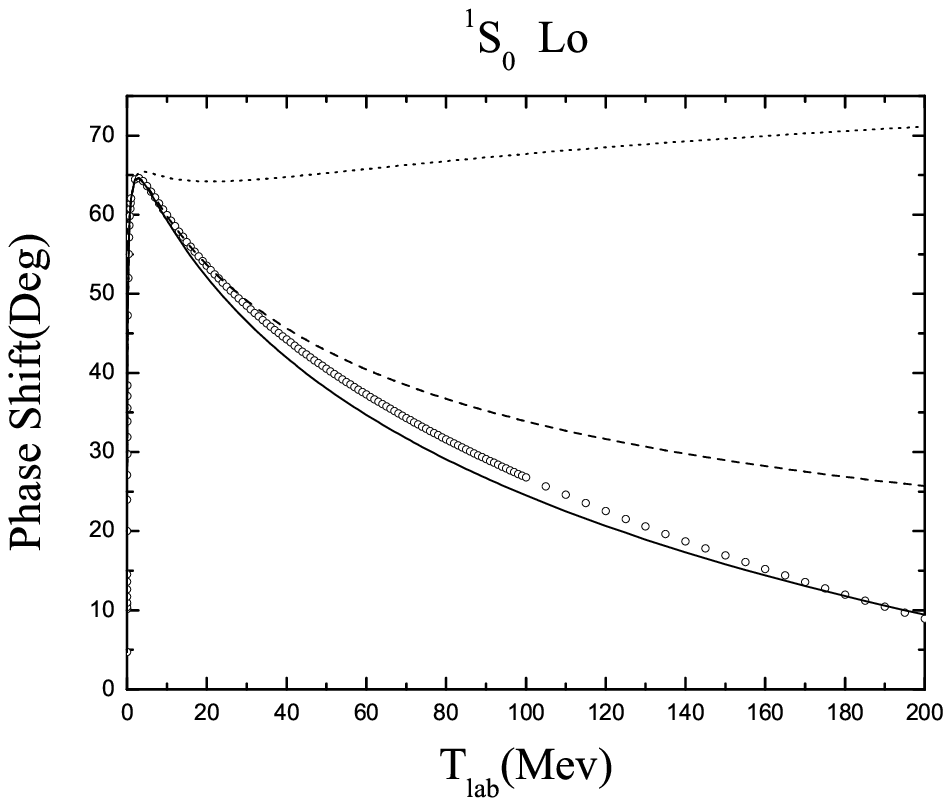}} &
    \resizebox{60mm}{!}{\includegraphics{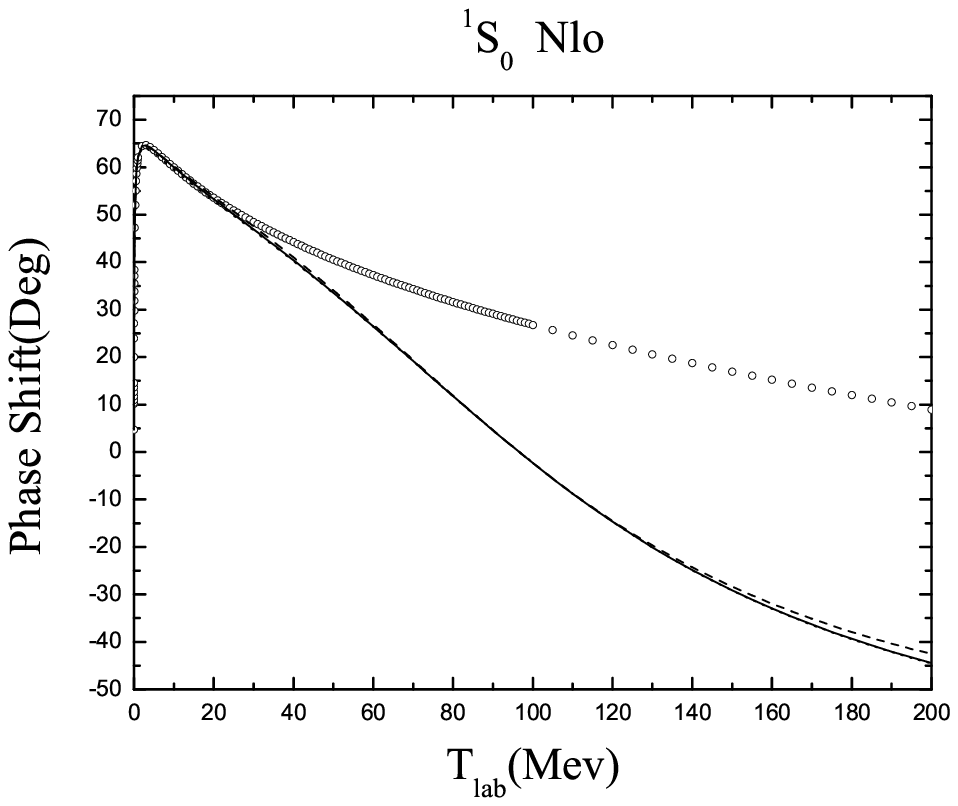}}&
    \resizebox{60mm}{!}{\includegraphics{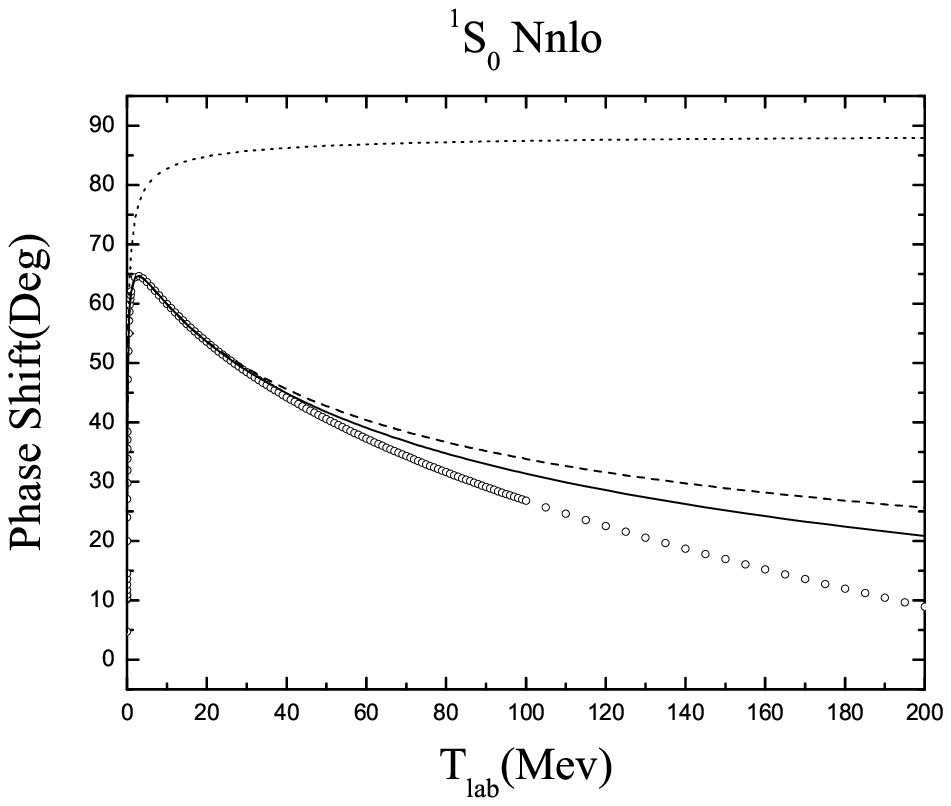}}
    \\ (a)&(b)&(c)
\end{tabular} \caption{\small Predictions for $^1S_0$ phase shifts
with three different Pad\'e approximants: (1)
$\text{Re}({\mathcal{G}})\approx g_{(0)}$ (dotted line); (2)
$\text{Re}({\mathcal{G}})\approx g_{(0)}+ g_{(1)} p^2$ (dashed
line); (3)
$\text{Re}({\mathcal{G}})\approx(1+\frac{\nu_1}{\nu_0}p^2)/(
\frac{\delta_0}{\nu_0}+\frac{\delta_1}{\nu_0}p^2)$ (solid line). The
circles denote the PWA data[23].  (a) for Lo; (b) for Nlo; (c) for
Nnlo .} \label{fixedpade}
\end{center}
\end{figure}

\begin{figure}[h]
\begin{center}
\begin{tabular}{ccc}
\hspace{-0.3cm} \resizebox{60mm}{!}{\includegraphics{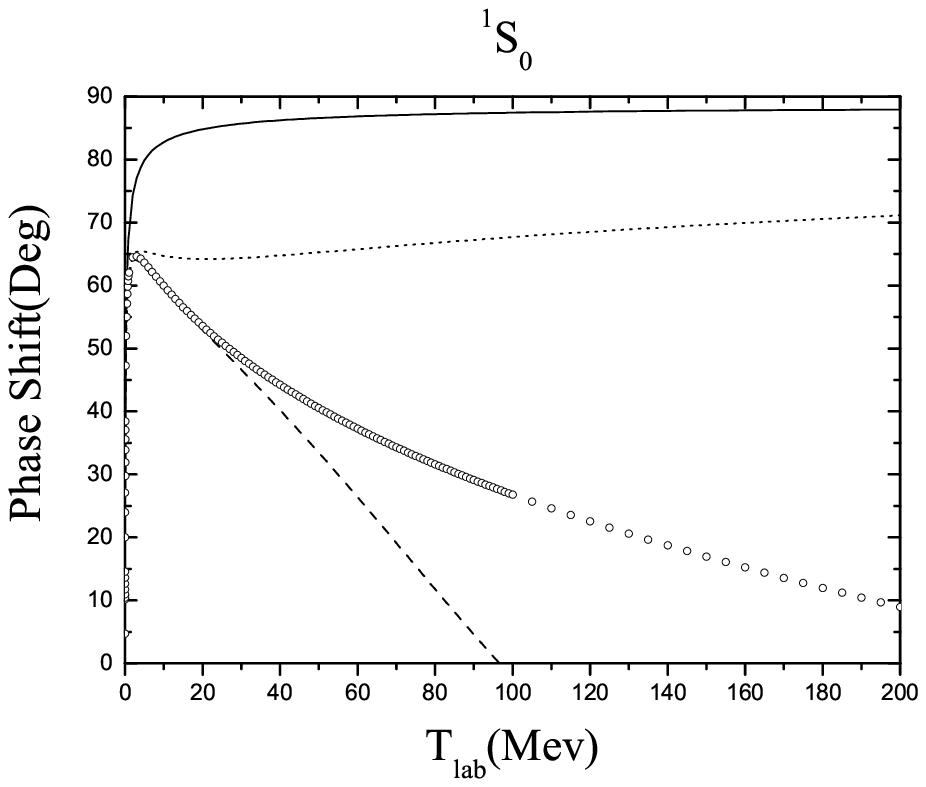}} &
\resizebox{60mm}{!}{\includegraphics{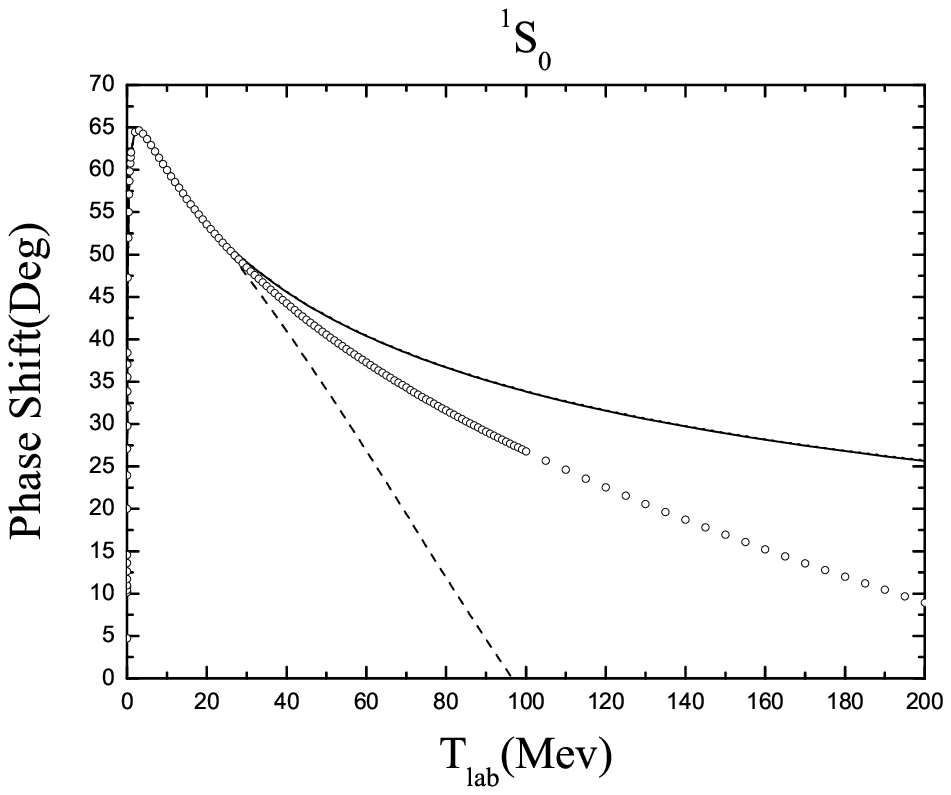}}&
    \resizebox{60mm}{!}{\includegraphics{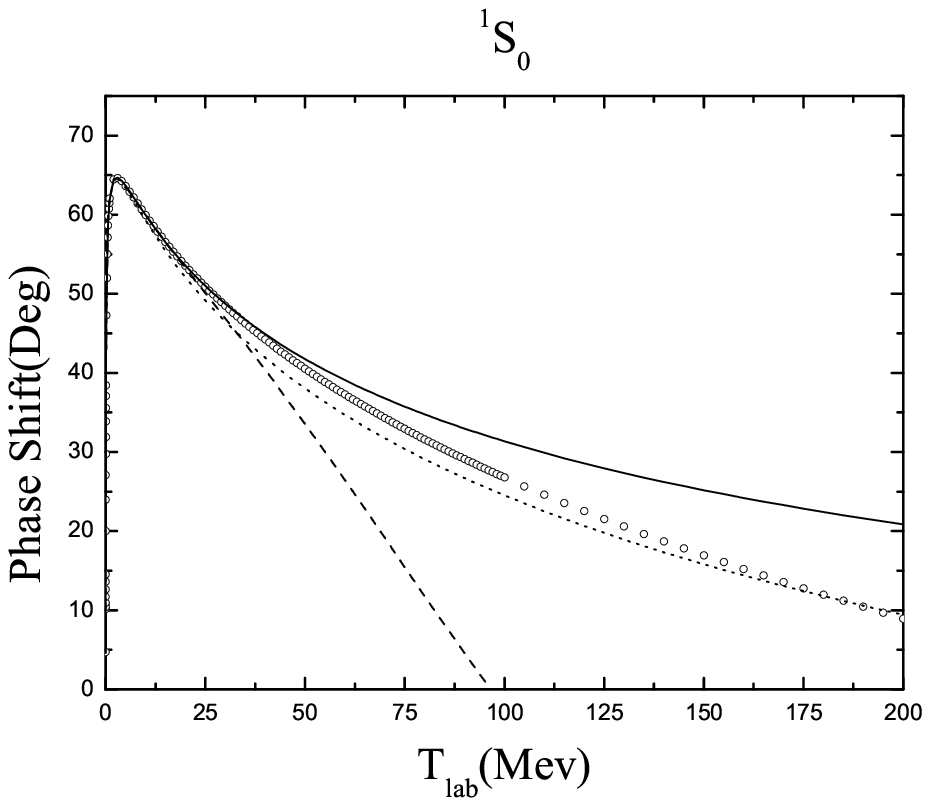}}
    \\ (a)&(b)&(c)
\end{tabular} \caption{\small Predictions for $^1S_0$ phase shifts at different
orders of potential: Dotted line for Lo, dashed line for Nlo and
solid line for Nnlo. The circles denote the PWA data[23]. (a) for
$\text{Re}({\mathcal{G}})\approx g_{(0)}$; (b) for
$\text{Re}({\mathcal{G}})\approx g_{(0)}+ g_{(1)} p^2$; (c) for
$\text{Re}({\mathcal{G}})\approx (1+\frac{\nu_1}{\nu_0}p^2)/(
\frac{\delta_0}{\nu_0}+\frac{\delta_1}{\nu_0}p^2)$ .}
\label{fixedorder}
\end{center}
\end{figure}

The phase shifts predicted at Lo, Nlo and Nnlo are depicted in
Fig.\ref{fixedpade} (a), (b), (c) respectively. At each order,
three Pad\'e approximants are shown respectively: (1)
$\text{Re}({\mathcal{G}})\approx g_{(0)}$ (dotted lines); (2)
$\text{Re}({\mathcal{G}})\approx g_{(0)}+g_{(1)} p^2$ (dashed
lines); (3) $\text{Re}({\mathcal{G}})\approx
(1+\frac{\nu_1}{\nu_0}p^2)/(
\frac{\delta_0}{\nu_0}+\frac{\delta_1}{\nu_0}p^2)$ (solid lines).
From these diagrams, one could find that, at each order, the
prediction improves as more Pad\'e parameters are present, which
is a natural tendency. One could also anticipate that with each
Pad\'e approximant, the predictions should also improve as higher
order terms are present in the potential, which are responsible
for the interactions at higher energy. The results are shown in
Fig.\ref{fixedorder}. In each figure, the predictions are compared
among different orders of potential with a fixed Pad\'e
approximant. Note that in Fig.\ref{fixedorder} (b), in the whole
range of the figure, the Lo curve almost identically coincides
with the Nnlo curve.

Globally, the improvement with chiral orders is obvious: Nnlo
prediction (solid line) is better than Nlo prediction (dashed
line), and Nlo prediction is better than Lo (dotted line). There
are also some interesting details: From these figures, we could
see that in the higher energy region, all the Nlo curves have
larger deviation from PWA data than the Nnlo curves. Some times,
they are even worse than the Lo curves (C.f. Fig.\ref{fixedorder}
(b) and (c)).

Here, we note that the solid curve in Fig.\ref{fixedpade} (a)
seems puzzling. With only leading order potential ($V_{1\pi}\sim
\frac{g_{\pi}^2}{f_{\pi}^2}\frac{\sigma_1\cdot q \sigma_2\cdot
q}{q^2+m^2_{\pi}}$ plus a contact term $V_c=C_0$), one obtains
pretty good predictions for the phase shifts, especially at the
higher energies. The reason lies in the Pad\'e approximant of
${\mathcal{G}}$,
$\text{Re}({\mathcal{G}})\approx(1+\frac{\nu_1}{\nu_0}p^2)/(
\frac{\delta_0}{\nu_0}+\frac{\delta_1}{\nu_0}p^2)$), which is
fixed 'physically' and effectively 'induces' higher order
interactions upon iteration. This result agrees with the findings
in Ref.\cite{Nieves}, where a nonperturbative of $T$-matrix is
obtained using a simple potential ($V=V_{1\pi}(\text{Lo})+V_{(2)}
(\text{in Eq.(\ref{Tn2}), Nlo})$, and the prediction of the phase
shifts is surprisingly good in a wider range of energy after the
nonperturbative divergences is removed through fitting. Our
analysis above provides a simple explanation of this surprise: The
nonperturbative renormalization is properly treated! A closer look
at the lower energy regions reveals that the higher order
predictions still dominate the lower order ones (Cf.
Fig.\ref{fixedorder} (c)). Generally, the lower order predictions
could 'win' at higher energies only by chance.

\section{Utilities of Pad\'e approximant and discussions} Now we
have seen that after the nonperturbative prescription dependence
is properly resolved (here realized through low energy region
fitting), the EFT approach facilitates 'physical' prediction for
low energy $NN$ scattering, at least in $^1S_0$ channel. Since no
specification of regularization and renormalization is needed, the
Pad\'e approximant of ${\mathcal{G}}$ defined in Eq.(\ref{npt}) in
fact provides a universal parametrization of the renormalization
prescription dependence of the $T$-matrix in nonperturbative
regimes. Note that both the potential and the Pad\'e approximant
of ${\mathcal{G}}$ could be systematically extended to higher
orders in EFT.

Comparing with previous results, we find that the Lo prediction
with $\text{Re}({\mathcal{G}})\approx(1+\frac{\nu_1}{\nu_0}p^2)/(
\frac{\delta_0}{\nu_0}+\frac{\delta_1}{\nu_0}p^2)$ (C.f,
Fig.\ref{fixedorder} (c)) differs significantly from that given in
Ref.\cite{Epel} and looks better. This nontrivial difference in
predictions at leading order reflects the importance of
nonperturbative renormalization. However, at higher chiral orders,
especially at Nnlo, our results show no obvious differences in
comparison with Ref.\cite{Epel}. That means, including higher
order interactions would lessen or tend to remove the nontrivial
nonperturbative renormalization prescription dependence. This is a
marvellous fact, since the fundamental requisite in EFT
application is that renormalization prescription dependence should
decrease as higher order interactions are included. Therefore, in
spite of being an approximation approach, the procedures described
above substantially proved or ascertained the rationality and
applicability of EFT method in nuclear forces in a very general
context. This is in sheer contrast to most known approaches where
a renormalization prescription must be specified and hence the
exploration of the prescription dependence is apparently limited.

So far we determined the Pad\'e parameters through fitting with
the potential defined by EGM\cite{Epel} where the contact
couplings were determined within a special cutoff scheme. In
principle, we should determine the couplings in a way that is more
prescription-independent: fitting through the combined space
$[C_{\ldots}]\bigcup[q_{\ldots},\mu]$, which should lead to a
better way for determining the EFT couplings. Now the Pad\'e
approximant provides us a convenient approach to do so without
really carrying out the formidable task of loop integrations and
renormalization to all orders. We will perform the investigations
along this line in the future. We believe other utilities could be
derived from this analysis and the parametrization Eq.(\ref{npt}).

From the above results, it is also obvious that it is fairly
sufficient to employ Pad\'e up to $\text{Re}({\mathcal{G}})\approx
(1+\frac{\nu_1}{\nu_0}p^2)/(
\frac{\delta_0}{\nu_0}+\frac{\delta_1}{\nu_0}p^2)$. For some
channels, say $P$-wave, it is often sufficient to use
$\text{Re}({\mathcal{G}})\approx g_{(0)}+ g_{(1)} p^2$, which will
be demonstrated in a separate report.

\section{summary}
In summary, we performed a Pad\'e analysis on a compact factor of
the $T$-matrix for $NN$ scattering so that the nonperturbative
prescription dependence and related effects could be conveniently
explored. Such analysis suggests a useful theoretical instrument
as well as a general and prescription-independent approach to test
the efficiency and rationality of the application of EFT methods
in nonperturbative regimes. Some related literature were also
explained and discussed in favor of our analysis.

\section*{Acknowledgement}
The authors are grateful to Dr. E. Epelbaum for helpful
communications. JFY is grateful to Professor B. A. Kniehl for his
hospitality at the II. Institute for Theoretical Physics of
Hamburg University where part of this work was done. This project
is supported in part by the National Natural Science Foundation
under Grant No. 10205004.

\vspace{3.0cm}

\begin{references}
\bibitem{WeinEFT}S. Weinberg, Phys. Lett.
{\bf B 251}, 288 (1990); Nucl. Phys. {\bf B 363}, 1 (1991).
\bibitem{BevK} See, e.g., P. Bedaque and U. van
Kolck, Ann. Rev. Nucl. Part. Sci. \textbf{52}, 339 (2002)
[nucl-th/0203055]; Ulf-G. Meissner, nucl-th/0409028.

\bibitem{vK}C. Ord\'o\~nez, L. Ray and U. van Kolck,
Phys. Rev. \textbf{C53}, 2086 (1996); U. van Kolck, Nucl. Phys.
\textbf{A645}, 327 (1999).
\bibitem{Epel}E. Epelbaum, W. Gl\"ockle and U. Meissner,
Nucl. Phys. \textbf{A671}, 295 (2000), Eur. Phys. J. \textbf{A15},
543 (2002), \textbf{A19}, 125,401 (2004).
\bibitem{Steele}J.V. Steele and R.J. Furnstahl, Nucl. Phys.
\textbf{A637}, 46 (1999).
\bibitem{Rho}T.S. Park, K. Kubodera,
D.P. Min and M. Rho, Phys. Rev. \textbf{C58}, R637 (1998).
\bibitem{KSW}D.B. Kaplan, M.J. Savage and M.B. Wise,
Phys. Lett. \textbf{B424}, 390 (1998); Nucl. Phys. \textbf{B534},
329 (1998); S. Fleming, T. Mehen and I.W. Stewart, Nucl. Phys.
\textbf{A677}, 313 (2000); Phys. Rev. \textbf{C61}, 044005 (2000).
\bibitem{Fred} T. Frederico, V.S. Tim\'oteo and L. Tomio, Nucl.
Phys. \textbf{A653}, 209 (1999).
\bibitem{Gege}J. Gegelia, Phys. Lett. \textbf{B463}, 133 (1999);
J. Gegelia and G. Japaridze, Phys. Lett. \textbf{B517}, 476
(2001); J. Gegelia and S. Scherer, nucl-th/0403052.
\bibitem{Soto}D. Eiras and J. Soto, Eur. Phys.
J. \textbf{A17}, 89 (2003)[nucl-th/0107009].
\bibitem{BBSvK}S.R. Beane, P. Bedaque, M.J. Savage and U. van Kolck,
Nucl. Phys. \textbf{A700}, 377 (2002).
\bibitem{RGE} T. Barford and M.C. Birse, Phys. Rev.
\textbf{C67}, 064006 (2003).
\bibitem{Oller}J.A. Oller, arXiv:nucl-th/0207086.
\bibitem{Nieves}J. Nieves, Phys. Lett. \textbf{B568}, 109
(2003)[nucl-th/0301080].
\bibitem{VA} M.P. Valderrama and E. Ruiz Arriola, Phys. Lett.
\textbf{B580}, 149 (2004); Phys. Rev.\textbf{C70}, 044006 (2004).
\bibitem{Higa} R. Higa, nucl-th/0411046.
\bibitem{chLnforce}S.R. Beane and M.J. Savage, Nucl. Phys.
\textbf{A713}, 148 (2003); Nucl. Phys. \textbf{A717}, 91 (2003);
E. Epelbaum, W. Gl\"ockle and U.-G. Meissner, Nucl. Phys.
\textbf{A714}, 535 (2003).
\bibitem{g_npt}J.-F. Yang, nucl-th/0310048v6(nucl-th/0407090).
\bibitem{contact} J.-F. Yang and J.-H. Huang, Phys. Rev.
\textbf{C71}, 034001(2005), {\em ibid.}, \textbf{C71}, 069901(E)
(2005)[nucl-th/0409023v3].
\bibitem{foot}The generalization to the coupled channels is
straightforward. We will perform such analysis in the coupled
channels in the future.
\bibitem{newton} R.G. Newton, {\em Scattering Theory of Waves and
Particles}, 2nd Edition (Springer-Verlag, New York, 1982), p187.

\bibitem{scheme}P.M. Stevenson, Phys. Rev. \textbf{D23}, 2916 (1981);
G. Grunberg, Phys. Rev. \textbf{D29}, 2315 (1984); S.J. Brodsky,
G. P. Lepage and P. B. Mackenzie, Phys. Rev. \textbf{D28}, 228
(1983).

\bibitem{nij}V.G.J. Stoks, R.A.M. Klomp, M.C.M. Rentmeester, and
J.J. de Swart, Phys. Rev. \textbf{C48}, 792 (1993).


\end{references}
\end{document}